\newif\ifshortver
\shortverfalse

\ifshortver
\documentclass[conference,a4paper]{IEEEtran}
\IEEEoverridecommandlockouts
\else
\documentclass[english,onecolumn]{IEEEtran}
\fi

\newcommand{\ackStmt}{
The work of Cheuk Ting Li was supported in part by the Hong Kong Research Grant Council Grant ECS No. CUHK 24205621.
The work of Sherman Chow was supported in part by the Hong Kong Research Grant Council Grant GRF Nos. CUHK 14209918 and 14210621.
The authors would like to thank the anonymous reviewers for their valuable comments.
\ifshortver
Some proofs are deferred to the full version~\cite{ace_arxiv} due to space constraints.
\fi
}

\ifshortver
\fi

\usepackage[utf8]{inputenc} 
\usepackage[T1]{fontenc}
\usepackage{url}              
\usepackage{cite}             

\usepackage[cmex10]{amsmath}  
\usepackage{mleftright}       
\mleftright                   

\usepackage{graphicx}         
\usepackage{booktabs}         

\usepackage{amssymb}
\usepackage{esint}
\makeatletter

\def\thmFlag{0}
\ifnum\thmFlag=1
\usepackage{amsthm}
\theoremstyle{plain}
\newtheorem{thm}{\protect\theoremname}
\theoremstyle{definition}
\newtheorem{defn}[thm]{\protect\definitionname}
\theoremstyle{plain}
\newtheorem{prop}[thm]{\protect\propositionname}
\theoremstyle{plain}
\newtheorem{lem}[thm]{\protect\lemmaname}
\theoremstyle{plain}
\newtheorem{cor}[thm]{\protect\corollaryname}
\theoremstyle{definition}
\newtheorem{example}[thm]{\protect\examplename}
\theoremstyle{definition}
\newtheorem{rem}[thm]{\protect\remarkname}
\else
\newtheorem{thm}{\protect\theoremname}
\newtheorem{defn}[thm]{\protect\definitionname}

\fi

\usepackage{mathrsfs}
\usepackage{algorithm,algpseudocode}

\usepackage{colortbl}
\definecolor{lightgray}{rgb}{0.9,0.9,0.9}
\definecolor{lightred}{rgb}{1,0.8,0.8}
\definecolor{lightgreen}{rgb}{0.6,1,0.6}
\definecolor{lightyellow}{rgb}{1,1,0.5}
\definecolor{lightgrey}{rgb}{0.8,0.8,0.8}

\allowdisplaybreaks[1]

\makeatother

\usepackage{babel}
\providecommand{\corollaryname}{Corollary}
\providecommand{\definitionname}{Definition}
\providecommand{\lemmaname}{Lemma}
\providecommand{\propositionname}{Proposition}
\providecommand{\theoremname}{Theorem}
\providecommand{\examplename}{Example}
\providecommand{\remarkname}{Remark}

\begin{document}

\title{Unconditionally Secure Access Control Encryption}

\author{%
  \IEEEauthorblockN{Cheuk Ting Li and Sherman S. M. Chow}
\ifshortver
\else
\\
\fi
  \IEEEauthorblockA{%
    Department of Information Engineering, The Chinese University of Hong
Kong\\
Email: \{ctli,smchow\}@ie.cuhk.edu.hk}
}


\maketitle
\begin{abstract}
Access control encryption (ACE) enforces, through a sanitizer as the mediator, that only legitimate sender-receiver pairs can communicate, without the sanitizer knowing the communication metadata, including its sender and recipient identity, the policy over them, and the underlying plaintext. Any illegitimate transmission is indistinguishable from pure noise. Existing works focused on computational security and require trapdoor functions and possibly other heavyweight primitives. We present the first ACE scheme with information-theoretic security (unconditionally against unbounded adversaries). Our novel randomization techniques over matrices realize sanitization (traditionally via homomorphism over a fixed randomness space) such that the secret message in the hidden message subspace remains intact if and only if there is no illegitimate transmission.
\end{abstract}




\section{Introduction}
Sensitive information should be safeguarded from arbitrary access.
Whether it is authorized depends on an access control policy, which assigns roles to users and determines whom they can read from or write to.
For example, consider the Bell--Lapadula (security/access-control) model~\cite{bell1973secure}, where users are divided into different clearance levels.
Transmissions are only allowed from a lower level (e.g., the public) to a higher level (e.g., the CEO).
Transmissions from higher to lower levels are forbidden to prevent the leakage of sensitive information.

Secure communication uses encryption, but it often only controls who can read but not who can write.
Signatures (or message authentication codes) also do not suffice.
When the data falls into the wrong hands, attackers probably do not care if they are signed.
Traditional access control requires a gatekeeper that sits between the users to check all requests against the policy before allowing the traffic to go through.
This puts a high level of trust in its well-behavior
and does not match the modern setting where communication is encrypted.

\emph{Meta-data}, i.e., the sender and recipient information, should also be protected.
It is unclear how a gatekeeper could enforce access control even with the meta-data.
A ciphertext could form a ``subliminal channel'' and become ``decryptable'' by an illegitimate receiver. 
For example, a pair of malicious sender (\emph{leaker}) and receiver (\emph{listener}) could agree apriori that the first bit of the ciphertext is the bit to convey, which can be obtained by, say, rejection sampling.

Cryptographic approaches enable ``blindfolded'' sanitization -- without leaking the meta-data, access control can still be enforced, sanitizing any subliminal channel.
This is formalized as access control encryption (ACE) by Damg{\aa}rd, Haagh, and Orlandi~\cite{damgaard2016access}.
Their setting considers all traffic to be encrypted and must go through a \emph{sanitizer} via an anonymous channel.
The sanitizer then uses a sanitization key to sanitize them without knowing their meta-data.
The sanitized ciphertext will then be safe to be broadcast to all receivers.

\medskip
\noindent
\emph{Technical Overview.}
We first go through the design of the first computationally secure linear-size ACE~\cite{damgaard2016access},
starting with the scheme for a single sender-receiver pair built from public-key encryption (PKE).
PKE allows the sender to send messages to the receiver with confidentiality, hence enforcing the ``no-read rule,'' which disallows illegitimate reading.
However, as argued (like~\cite{damgaard2016access}), PKE alone does not forbid illegitimate writing from any sender/leaker to any receiver/listener.
To enforce such a ``no-write rule,'' the sanitizer would apply a randomized transformation on the ciphertext such that in case of illegitimate communication, a sanitized ciphertext looks like a random string in the ciphertext space, which encodes no information.
Meanwhile, to avoid interfering with legitimate transmissions, the transformation exploits the homomorphic properties of PKE.\footnote{(Many-time-secure) PKE features probabilistic encryption $\mathsf{Enc}(m; r)$, where $m$ is from a message space, and $r$ is from a randomness space.
(Linearly-)homomorphic PKE satisfies $\mathsf{Enc}(m_1; r_1) + \mathsf{Enc}(m_2; r_2) = \mathsf{Enc}(m_1 + m_2; r_1 + r_2)$ (and $s\cdot\mathsf{Enc}(m; r)= \mathsf{Enc}(sm; sr)$ for random~$s$).}
In short, homomorphism crucially enables the \emph{sanitization logic} that only interferes with message $m$ when the policy disallows, \emph{without} knowing $m$, the metadata (who the sender and receiver are), or an explicit policy check.


For a general policy over $n$ roles, their scheme~\cite{damgaard2016access} runs $n$ single-pair instances.
Each receiver $b$ is given the decryption key 
for the $b$-th instance.
Each sender $a$ who can legitimately write to receivers $b$'s is given encryption keys for those instances.
For each of the non-receivers, the sender samples a random string indistinguishable from valid encryption.
The resulting ciphertext is of size linear in the number of roles.


\medskip
\noindent
\emph{Challenges.}
Several mismatches make it unclear how to port this design to the information-theoretic (IT) setting and pose the challenges this paper is going to solve.
We summarize them as two core functionalities.
1) Public-key primitives are used (despite encryption requiring a secret key), which introduce asymmetry between the sender and receiver roles and readily attain collusion resistance -- collusion between any sender and any receiver cannot violate the no-write rule.
2) More importantly, their homomorphisms 
enable the aforementioned sanitization logic for processing encrypted metadata.
PKE is not immediately available in the IT setting.
IT-secure symmetric key encryption schemes, such as the one-time pad, do not come with any enforcement of asymmetric access-control policies such as the Bell--Lapadula model~\cite{bell1973secure}.
If a sender can send to a recipient, then the recipient can send to the sender as well since they share the same key.
Section~\ref{sec:basic} illustrates several failed attempts and details the corresponding (collusion) attacks before introducing our IT-secure scheme.

There are also structural differences.
Traditional schemes
(e.g., the linear-size scheme~\cite{damgaard2016access} and constant-size schemes~\cite{wang2021cross}) 
explicitly separate the message and randomness spaces.
Our scheme features a singleton message space and (a singleton) ciphertext space that a leaker could exploit fully.
Recall that any ciphertext, legitimate or not, should look uniform to a sanitizer.
In other words, the sanitization logic needs to ensure ``somehow'' that the \emph{secret message} in the \emph{hidden} message subspace remains intact if and only if there is no illegitimate transmission, 
which we realize by our proposed randomization techniques over matrices, deviating from traditional tricks.

Our techniques allow an intuitive ACE construction --
The sender applies a random linear transformation 
(given by the encryption key) 
to place the message into a random subspace of the ciphertext space.
The sanitizer then applies another random linear transformation 
(stems from the sanitization key) on the ciphertext.
Finally, the recipient applies a suitable linear transformation (the decryption key) to undo the previous two transformations.
We exploit the non-commutative nature of matrix multiplication to realize encryption functionalities
and formally assert (in Section~\ref{sec:basic}) that this technique avoids the weaknesses of symmetric-key encryption mentioned above.



\medskip
\noindent
\emph{Summary of Contribution.}
\begin{itemize}
\item
We formalize 
ACE in the IT setting.
Beyond the perfect no-read rule,
we propose an IT variant of the no-write rule,
capturing a new kind of adversary yet to be covered by existing counterparts in the computational setting.
\item
We propose the first 
ACE scheme in the IT setting with provable security against collusions.
\item
Our random matrix technique is of independent interest and might find applications elsewhere.
\end{itemize}

\medskip
\noindent
\emph{Related Works.}
The seminal work also proposes a scheme with polylogarithmic complexity~\cite{damgaard2016access} based on indistinguishability obfuscation.
Kim and Wu~\cite{kim2017access} proposed another construction from functional encryption.
Sanitization in both schemes uses an ``encrypted circuit'' to create a new ciphertext out of the original one, which is a heavyweight approach in general.

Wang and Chow~\cite{wang2021cross} propose the first general-policy ACE scheme that features constant-size ciphertexts with practical efficiency (without general-purpose circuit-based primitives).
Wang, Wong, and Chow~\cite{wang2021access} construct (lattice-based) ACE from 
group encryption, which is additionally post-quantum secure.
In particular, the ciphertext size is still constant.\footnote{Constant size 
here refers to the number of cryptographic group elements, following the convention of (computationally secure) cryptographic schemes (treating the security parameter as a constant).
In both schemes, the encryption key size is constant, while the decryption key size is linear in the maximum number of senders that a user can receive from according to the policy.
Note that the linear-size scheme~\cite{damgaard2016access} requires an encryption key to be of size linear in the maximum number of receivers that a user can legitimately send to.}
Finally, some schemes support specific policy families~\cite{fuchsbauer2017access,sedaghatP2021cross}.
All these works are only secure in the computational setting.

An existing information-theoretic access control work~\cite{ferrara2003information} proposes a key assignment scheme to ensure that only a subset of parties can access each piece of information.
It does not forbid illegitimate sender-receiver pairs from communicating.

Many IT-secure schemes, e.g., secret sharing~\cite{cacm/Shamir79,massey1993minimal,yuan2005secret} and secure network coding~\cite{cai2002secure,feldman2004capacity}, use linear transformations over finite fields.
We use 
linear transformations 
as secret keys, 
sharing some similarities with the classical Hill cipher~\cite{hill1929cryptography}.

\medskip
\noindent
\emph{Notations.}
Write $[a..b] = \{a, a + 1, \ldots, b\}$, $[n] = [1..n]$.
The probability mass function (PMF) of a random variable $X$ is $p_X$, and the conditional PMF of $X$ given $Y$ is $p_{X|Y}$.
For two PMFs $p, q$, their statistical distance (or total variation distance) is $\mathrm{SD}(p,q)=(1/2)\sum_x |p(x)-q(x)|$.

\medskip
\section{Problem Setting}

Consider $n$ parties who can communicate anonymously via a sanitizer,
and a directed graph $E$ with vertex set $[n]$ (the security policy) that describes which pairs of parties are allowed to communicate.
First, the central authority (CA) distributes private keys to the parties and the sanitizer.\footnote{Wang and Chow~\cite{wang2021cross} first consider the cross-domain setting, in which there is a sender authority and a receiver authority instead of a single CA.
Their scheme also allows keyless sanitization~\cite{wang2021access}, unlike earlier works~\cite{damgaard2016access,kim2017access}.}
Next, one of the parties, party $a$, wishes to send a message $M$ (independent of the keys) to party $b$ where $(a, b)\in E$.
To do so, party $a$ sends a ciphertext
$C$ anonymously to the sanitizer (i.e., the identity of $a$ is not revealed to the sanitizer).
The sanitizer then sanitizes $C$ into $C'$ and broadcasts~$C'$ to all parties.
Party~$b$ should recover $M$ using~$C'$.
On the other hand, if $(a,b)\notin E$, then party $a$ must be disallowed to send messages to party $b$.
For example, the Bell--Lapadula policy~\cite{bell1973secure} corresponds to $E=\{(i,j):i<j\}$, where a party can only send messages to someone at a higher~level.

More precisely, 
adapting the seminal Damg{\aa}rd--Haagh--Orlandi formulation~\cite{damgaard2016access}
into an information-theoretic security setting (without bounds on the computational
power of attackers),
an ACE scheme over the policy $E$, message space $\mathcal{M}$, ciphertext space $\mathcal{C}$, sanitized ciphertext space $\mathcal{C}'$, with the distance parameter $\epsilon > 0$, consists of the following algorithms:
\begin{itemize}
\item \textbf{Key generation:} $(K_{0},K_{1},\ldots,K_{n})\leftarrow\mathsf{KeyGen}()$.
The CA outputs the sanitizer key $K_{0}$ and the private
keys $K_{1},\ldots,K_{n}$, where $K_{i}$ is for party $i$.
\item \textbf{Encryption:} $C\leftarrow\mathsf{Enc}(a,b,K_{a},M)$.
If party $a$ wishes to send a message $M$ to party $b$ where $(a,b)\in E$,
$\mathsf{Enc}$ encrypts $M\in\mathcal{M}$ into ciphertext $C \in\mathcal{C}$ using key $K_{a}$.
\item \textbf{Sanitization:} $C'\leftarrow\mathsf{San}(K_{0},C)$.
The sanitizer uses the sanitization key $K_{0}$ to sanitize $C$ into $C' \in\mathcal{C}'$,
which will be broadcast.
Note that the sanitizer does not observe
the identities of $a,b$, nor whether $(a,b)\in E$.
\item \textbf{Decryption:} $\hat{M}\leftarrow\mathsf{Dec}(a,b,K_{b},C')$.
Party $b$, who knows the identity of the sender $a$, tries to recover $M$ as $\hat{M}$ using $C'$ and the key $K_{b}$.
\end{itemize}

The scheme must satisfy the following three requirements:

\medskip
\begin{defn}[Correctness]
$M=\hat{M}$ if $(a,b)\in E$.
\end{defn}
\medskip

\medskip
\begin{defn}[Perfect no-read rule]
The conditional distribution $p_{C|(K_{i})_{i\in[0..n] \backslash \{a,b\}},M}(\,\cdot\, | (k_{i})_{i\in[0..n]\backslash\{a,b\}},m)$ is $\mathrm{Unif}(\mathcal{C})$, regardless of the values of 
$(a,b,(k_{i})_{i\in[0..n]\backslash\{a,b\}},m)$.
\end{defn}
\medskip

To an attacker which knows a subset of the keys $K_A = (K_i)_{i \in A}$ (which may include the sanitizer key), where $A \subseteq [0..n] \backslash \{a, b\}$,
the conditional distribution of $C$ given $M$ according to its knowledge 
is $p_{C | K_A, M} = \mathrm{Unif}(\mathcal{C})$ by the no-read rule.
This implies $C \perp (K_A, M)$ ($C$ is independent of $(K_A, M)$), which, combined with $K_A \perp M$, implies that $M \perp (K_A, C)$, i.e., the attacker cannot gain any knowledge of~$M$ using $C$.
Moreover, the conditional distribution $p_{C | K_A, M}$ does not depend on $a,b$ as long as $a,b \notin A$, meaning that $C$ provides no additional information about $a,b$ (other than that $a,b \notin A$). The identities of the sender $a$ and the receiver $b$ are kept secret from the sanitizer and all other parties.



\medskip
\begin{defn}[No-write rule]
Consider a leaker which has access to a subset $K_{A}=(K_{a})_{a\in A}$ of the keys of parties $A\subseteq[n]$, 
and a listener which has access to another subset $K_{B}$, $B\subseteq[n]$, where $(a,b)\notin E, \forall a\in A$, $b\in B$,
i.e., the leaker cannot write to the listener legitimately.
The leaker sends $\tilde{C}$:
\[
\tilde{C} \leftarrow \mathsf{Att}(A,B,K_{A}),\text{~where~}\tilde{C} \in \mathcal{C}
\]
to the sanitizer, which produces 
$\tilde{C}'\leftarrow\mathsf{San}(K_{0},\tilde{C})$.
We require 
\begin{equation}\label{eq:setting_sd}
\mathrm{SD}\big(p_{\tilde{C}',K_{B}},\,\mathrm{Unif}(\mathcal{C}') \times p_{K_{B}}\big) \le \epsilon
\end{equation}
for any $A, B$, and $\mathsf{Att}$,
i.e., the statistical distance (or total variation distance) between the joint distribution of $\tilde{C}',K_{B}$ and their ideal joint distribution (where $\tilde{C}'$ is uniform and independent of $K_{B}$) is no larger than the distance parameter~$\epsilon$.
According to the listener
(which knows $K_{B}$), $\tilde{C}'$ is almost
indistinguishable from uniform noise.
\end{defn}
\medskip


Our definition gives the leaker and listener each a set of keys to capture collusion in the real world.
$\mathsf{Att}$ captures whatever the leaker can send
without compromising the sanitizer.
The no-write rule forbids the leaker from sending to the listener as long as $(a,b) \notin E, \forall a \in A$, $b \in B$.
Notably, $A\cap B=\emptyset$ is not required.

Note that the no-write rule does not explicitly involve a message that the leaker is trying to send to the listener.
We now show that the no-write rule implies that the leaker cannot send such a message. Loosely speaking, the no-write rule ensures that the ``channel'' from the leaker to the listener has an output that is almost indistinguishable from uniform noise regardless of the channel input, and hence such channel cannot be used to convey information.
Consider 
the leaker encodes a random message $M$ with a known probability mass function $p_M$
into $\tilde{C} \leftarrow \mathsf{Att}_M(A, B, K_{A})$.
The listener runs a decoding algorithm $\mathsf{AttDec}(A, B, K_{B},\tilde{C}')$ 
to recover $\hat{M}$
after observing 
$\tilde{C}' \leftarrow \mathsf{San}(K_{0},\tilde{C})$.
By the no-write rule \eqref{eq:setting_sd}, we know that
$\mathrm{SD}(p_{\tilde{C}',K_{B}|M=m},\,\mathrm{Unif}(\mathcal{C}')\times p_{K_{B}})\le\epsilon$
for each~$m$.
So, $\mathrm{SD}(p_{\tilde{C}',K_{B},M},\,\mathrm{Unif}(\mathcal{C}')\times p_{K_{B}}\times p_M)\le\epsilon$.
In the hypothetical situation where $\tilde{C}',K_{B},M$ are distributed according to $\mathrm{Unif}(\mathcal{C}')\times p_{K_{B}}\times p_M$, the probability of correct guess is bounded by $\mathbb{P}(\hat{M}=M) \le \max_{m} p_M(m)$, i.e., the listener can only guess the most probable $m$ since $M$ is independent of $(\tilde{C}',K_{B})$. Hence, in the actual situation, which is within $\epsilon$ in statistical distance from the hypothetical situation, we have
\begin{align*}
 & \mathbb{P}(\hat{M}=M) \le \max_{m} p_M(m) + \epsilon,
\end{align*}
i.e., the listener's guess of the message $M$ cannot be much better than the guess of someone who observes nothing other than $p_M$ (whose optimal guess is $\mathrm{argmax}_m p_M(m)$).




\medskip
\noindent
\emph{Comparisons.}
Computational formulations~\cite{damgaard2016access} grant attackers access to the encryption oracle $\mathcal{O}_{\mathsf{E}}$ that produces a sanitized ciphertext when given a message and the sender identity.
In the IT setting, unlimited oracle accesses by a computationally unbounded attacker would result in the attacker knowing all relevant information out of the oracle.
Such oracle access is thus uncommon among information-theoretic security settings.

For aspects where the no-write rule in this paper is stronger, we consider two separate attackers: a leaker and a listener.
They are given different but possibly overlapping sets of keys.
For example, consider $n = 2$, $E = {(1, 2)}$, 
i.e., party~2 can read from party 1.
The following scenarios are covered:
\begin{enumerate}
 \item the leaker has $\{K_2\}$, while the listener has $\{K_1, K_2\}$;
 \item the leaker has $\{K_1, K_2\}$, while the listener has $\{K_1\}$.
\end{enumerate}
The leaker in Case 1) has no keys with any writing power.
In Case 2), the leaker can write, but the listener cannot read.
Intuitively, the no-write rule should remain enforceable.


In contrast, traditional definitions~\cite{damgaard2016access} consider only a single (global) two-phase attacker, 
which can only choose one of $K_1$ or $K_2$, since accessing both will trivially break the security.
Interestingly, the recent formulation of Wang and Chow~\cite{wang2021cross} strengthens the original no-write rule~\cite{damgaard2016access} toward capturing Case 1).
Namely, the attacker can access additional encryption keys (e.g., $K_1$ in the above example) after seeing the challenge ciphertext (which can originate from $K_1$ as well).

\medskip
\section{Our Random Matrix Scheme for A User Pair}\label{sec:basic}

Before we present the scheme for general access policies in Section~\ref{sec:general}, 
we consider its building block, which works for the case $n = 2$ and $E = {(1, 2)}$, i.e., we only allow party~1 to send to party 2.
Party 1 encrypts $M$ into $C$ using $K_{1}$.
The sanitizer further encrypts $C$ into $C'$ using $K_{0}$.
Intuitively, party~2 should be able to decrypt both
layers using~$K_{2}$.

We first describe attempts that do not work to illustrate our final design incrementally.
If we use the one-time pad for both
ciphers, taking $M, K_{1}, K_{2} \in \{0, 1\}^{L}$,
$K_{0} = K_{1} \oplus K_{2}$,
then we can have $C \leftarrow M \oplus K_{1}$,
$C' \leftarrow C\oplus K_{0}$,
$\hat{M} \leftarrow C'\oplus K_{2}$.
While the perfect no-read rule
is satisfied, the no-write rule is violated since party 2 can send
a message $\tilde{M}\in\{0,1\}^{L}$ to party 1 by $\tilde{C}\leftarrow M\oplus K_{2}$,
$\tilde{C}'\leftarrow\tilde{C}\oplus K_{0}$, $\hat{\tilde{M}}\leftarrow\tilde{C}'\oplus K_{1}$.

This failed attempt shows that encryption should be non-commutative, so party 2 cannot cancel out the sanitization.
An example of non-commutative operations is matrix multiplications.
Consider a finite field~$\mathbb{F}$.
Taking $M=\mathbf{m}\in\mathbb{F}^{L} \backslash\{\mathbf{0}\}$,
$K_{1} = \mathbf{K}_{\mathrm{E}}\in\mathbb{F}^{L\times L}$, $K_{2} = \mathbf{K}_{\mathrm{D}}\in\mathbb{F}^{L\times L}$
as randomly chosen full-rank matrices, and $K_{0} = \mathbf{K}_{\mathrm{R}}=\mathbf{K}_{\mathrm{D}}^{-1}\mathbf{K}_{\mathrm{E}}^{-1}$,
we can have $\mathbf{c}\leftarrow\mathbf{K}_{\mathrm{E}}\mathbf{m} \in \mathbb{F}^{L} \backslash\{\mathbf{0}\}$,
$\mathbf{c}'\leftarrow\mathbf{K}_{\mathrm{R}}\mathbf{c} \in \mathbb{F}^{L} \backslash\{\mathbf{0}\}$, $\hat{\mathbf{m}}\leftarrow\mathbf{K}_{\mathrm{D}}\mathbf{c}'$.
There is no obvious way for party 2 to send messages to party 1, since
it cannot cancel out $\mathbf{K}_{\mathrm{D}}^{-1}$ by controlling
$\tilde{\mathbf{c}}$ in $\tilde{\mathbf{c}}'=\mathbf{K}_{\mathrm{R}}\tilde{\mathbf{c}}=\mathbf{K}_{\mathrm{D}}^{-1}\mathbf{K}_{\mathrm{E}}^{-1}\tilde{\mathbf{c}}$.

Nevertheless, the no-write rule is still violated since party 1 and party 2 can collude to obtain $\mathbf{K}_{\mathrm{R}}=\mathbf{K}_{\mathrm{D}}^{-1}\mathbf{K}_{\mathrm{E}}^{-1}$,
and send a message $\tilde{\mathbf{m}}$ to everyone by $\tilde{\mathbf{c}}\leftarrow\mathbf{K}_{\mathrm{R}}^{-1}\tilde{\mathbf{m}}$,
resulting in $\tilde{\mathbf{c}}'=\tilde{\mathbf{m}}$.
Therefore, it is also important that the sanitizer key cannot be obtained from
the encryption and decryption keys.

\medskip
\noindent
\emph{Random Matrix Scheme.}
Our scheme utilizes a sanitizer
key $\mathbf{K}_{\mathrm{R}}$ larger than $(\mathbf{K}_{\mathrm{E}}, \mathbf{K}_{\mathrm{D}})$ such that it cannot be obtained from $(\mathbf{K}_{\mathrm{E}}, \mathbf{K}_{\mathrm{D}})$.
Let $\mathbb{F}$ be a finite field of order~$q$.
Write 
\[
\mathrm{FR}(\mathbb{F}^{L\times N})=\left\{ \mathbf{A}\in\mathbb{F}^{L\times N}:\,\mathrm{rank}(\mathbf{A})=\min\{L,N\}\right\} 
\]
for the set of full-rank matrices.
To sample from
$\mathrm{FR}(\mathbb{F}^{L\times N})$ uniformly at random, we can repeat generating matrices from $\mathbb{F}^{L\times N}$ with i.i.d. entries until
we encounter a full rank matrix.\footnote{The probability that a random matrix in $\mathbb{F}^{L\times N}$ is full rank is $\prod_{i=|L-N|+1}^{\max\{L,N\}}(1-|\mathbb{F}|^{-i}) \ge 0.28$~\cite{blake2006properties}.}

Our proposed random matrix scheme is given as follows.
It has three parameters: the field size $q = |\mathbb{F}|$, the plaintext length $L\ge1$, and the ciphertext length
$N > 2L$, which collectively determine the distance parameter to be shown in Theorem \ref{thm:sec}.
\begin{itemize}
\item \textbf{Key generation:} Sample uniformly at random $(K_{0},K_{1},K_{2})=(\mathbf{K}_{\mathrm{R}},\mathbf{K}_{\mathrm{E}},\mathbf{K}_{\mathrm{D}})$
from the set
\begin{align*}
\mathcal{K}=\Big\{ & (\mathbf{A}_{\mathrm{R}},\mathbf{A}_{\mathrm{E}},\mathbf{A}_{\mathrm{D}})\in\mathrm{FR}(\mathbb{F}^{N\times N})\times\mathrm{FR}(\mathbb{F}^{N\times L})\\
 & \;\;\times\mathrm{FR}(\mathbb{F}^{L\times N}): \,\mathbf{A}_{\mathrm{D}}\mathbf{A}_{\mathrm{R}}\mathbf{A}_{\mathrm{E}}=\mathbf{I}_{L}\Big\}.
\end{align*}
One 
way
is to sample
$\mathbf{T}\sim \mathrm{Unif}(\mathbb{F}^{L\times(N-L)})$,
$\mathbf{S}_{\mathrm{E}},\mathbf{S}_{\mathrm{D}} \sim \mathrm{Unif}(\mathrm{FR}(\mathbb{F}^{N\times N}))$,
 independently, and set 
\begin{align}
\mathbf{K}_{\mathrm{E}}&
\leftarrow\mathbf{S}_{\mathrm{E}}\left[\begin{array}{c}
\mathbf{I}_{L}\\
\mathbf{0}^{(N-L)\times L}
\end{array}\right], 
\nonumber \\
\mathbf{K}_{\mathrm{D}}
&\leftarrow\left[\mathbf{I}_{L}\,\big|\,\mathbf{T}\right]\mathbf{S}_{\mathrm{D}},
\nonumber \\
\mathbf{K}_{\mathrm{R}}
&
\leftarrow\mathbf{S}_{\mathrm{D}}^{-1}\mathbf{S}_{\mathrm{E}}^{-1}.
\label{eq:keygen}
\end{align}
$\mathbf{K}_{\mathrm{R}},\mathbf{K}_{\mathrm{E}},\mathbf{K}_{\mathrm{D}}$
are clearly pairwise independent.
The explanation of why \eqref{eq:keygen} gives the uniform distribution over $\mathcal{K}$ will be given later.
\item \textbf{Encryption:} To encrypt message 
$\mathbf{m} \in \mathcal{M} = \mathbb{F}^{L} \backslash \{\mathbf{0}\}$,
party 1 sends 
$\mathbf{c} \leftarrow \mathbf{K}_{\mathrm{E}}\mathbf{m}$, where 
$\mathbf{c} \in \mathcal{C} = \mathbb{F}^{N} \backslash \{\mathbf{0}\}$.
\item \textbf{Sanitization:} Sanitizer outputs 
$\mathbf{c}' \leftarrow \mathbf{K}_{\mathrm{R}}\mathbf{c}$, where
$\mathbf{c}' \in \mathcal{C}' = \mathbb{F}^{N}\backslash\{\mathbf{0}\}$.
\item \textbf{Decryption:} Party 2 recovers 
$\hat{\mathbf{m}} \leftarrow \mathbf{K}_{\mathrm{D}}\mathbf{c}'$.
\end{itemize}

\medskip

To see why \eqref{eq:keygen} gives the uniform distribution over $\mathcal{K}$, note that if $(\mathbf{K}_{\mathrm{R}},\mathbf{K}_{\mathrm{E}},\mathbf{K}_{\mathrm{D}})$ is uniform over $\mathcal{K}$, then
$\mathbf{K}_{\mathrm{E}}$ is uniform over $\mathrm{FR}(\mathbb{F}^{N\times L})$ by symmetry. We can assume $\mathbf{K}_{\mathrm{E}} = \mathbf{S}_{\mathrm{E}}\left[\begin{array}{c}
\mathbf{I}_{L}\\
\mathbf{0}^{(N-L)\times L}
\end{array}\right]$ for a uniformly random $\mathbf{S}_{\mathrm{E}} \in \mathrm{FR}(\mathbb{F}^{N\times N})$. Conditional on $\mathbf{S}_{\mathrm{E}}$, we know that $\mathbf{K}_{\mathrm{R}}$ is uniform over $\mathrm{FR}(\mathbb{F}^{N\times N})$ by symmetry, so we can take $\mathbf{K}_{\mathrm{R}} = \mathbf{S}_{\mathrm{D}}^{-1}\mathbf{S}_{\mathrm{E}}^{-1}$ for a uniformly random $\mathbf{S}_{\mathrm{D}} \in \mathrm{FR}(\mathbb{F}^{N\times N})$. Conditional on $\mathbf{S}_{\mathrm{E}},\mathbf{S}_{\mathrm{D}}$, we know that $\mathbf{K}_{\mathrm{D}}$ is uniform over matrices that satisfy $\mathbf{K}_{\mathrm{D}}\mathbf{S}_{\mathrm{D}}^{-1}\left[\begin{array}{c}
\mathbf{I}_{L}\\
\mathbf{0}^{(N-L)\times L}
\end{array}\right] = \mathbf{I}_{L}$. Such $\mathbf{K}_{\mathrm{D}}$ must be in the form $\left[\mathbf{I}_{L}\big|\mathbf{T}\right]\mathbf{S}_{\mathrm{D}}$ for some $\mathbf{T} \in \mathbb{F}^{L\times(N-L)}$, giving the method in~\eqref{eq:keygen}.

\medskip

\noindent
\emph{Efficiency.}
We discuss the complexities of our 
scheme:
\begin{itemize}
\item Key lengths (in bits): $N^{2}\log q$ for sanitizer key $\mathbf{K}_{\mathrm{R}}$,
$NL\log q$ for each encryption/decryption key $\mathbf{K}_{\mathrm{E}}$/$\mathbf{K}_{\mathrm{D}}$.
\item Message length: $\log(q^{L}-1)\approx L\log q$.
\item Ciphertext (sanitized/unsanitized) length: $N\log q$.
\item Key generation time complexity: $O(N^3 (\log q)^2)$ for checking that a matrix in $\mathbb{F}^{N\times N}$ is full rank, and matrix multiplication
(which can be improved by faster matrix multiplication and finite field operation algorithms).
\item Encryption and decryption complexity: $O(LN(\log q)^{2})$.
\item Sanitization time complexity: $O(N^2(\log q)^{2})$.
\end{itemize}


The ratio between the sanitizer key length and the message length is $(N^{2}\log q)/(L\log q)$. With $N > 2L$, it must be 
at least $(2L+1)^{2}/L$, and attains 
the minimum of $9$ when $L = 1$, $N=3$.
Likewise, it can be checked that $L = 1$, $N=3$ also minimizes the ratio between the encryption/decryption key length and the message length, which is $3$.
Nevertheless, it does not minimize the ratio between the ciphertext length and the message length, which can be arbitrarily close to $2$.



\medskip
\noindent
\emph{Security.}
We now prove the main result, which shows that the random matrix scheme is secure.

\medskip
\begin{thm}\label{thm:sec}
The random matrix scheme for $|\mathbb{F}|=q$, $n=2$, $E=\{(1,2)\}$
attains a distance parameter~\eqref{eq:setting_sd}
\[
\epsilon= 2 q^{-(N/2-L)}.
\]
\end{thm}

\ifshortver
\begin{IEEEproof}[Proof of the first part]
\else
\begin{IEEEproof}
\fi
To check the perfect no-read rule, note that $\mathbf{c}=\mathbf{K}_{\mathrm{E}}\mathbf{m}$
is independent of $\mathbf{m}$, and $\mathbf{K}_{\mathrm{R}}$ is
independent of $\mathbf{K}_{\mathrm{E}}$, which follows from the
pairwise independence in \eqref{eq:keygen}.
Hence, $\mathbf{c},\mathbf{m},\mathbf{K}_{\mathrm{R}}$
are mutually independent, and the conditional distribution of $\mathbf{c}=\mathbf{K}_{\mathrm{E}}\mathbf{m}$
given $(\mathbf{m},\mathbf{K}_{\mathrm{R}})$ is always the uniform
distribution over $\mathbb{F}^{N}\backslash\{\mathbf{0}\}$ regardless
of $(\mathbf{m},\mathbf{K}_{\mathrm{R}})$.

Before proving the no-write rule, we first consider an alternative
method of generating $(\mathbf{K}_{\mathrm{R}},\mathbf{K}_{\mathrm{E}},\mathbf{K}_{\mathrm{D}})\leftarrow\mathcal{K}$.
Consider $\mathbf{K}_{\mathrm{R}},\mathbf{K}_{\mathrm{E}},\mathbf{K}_{\mathrm{D}}$
generated using \eqref{eq:keygen}.
Fix any $\mathbf{Z} \in \mathrm{FR}(\mathbb{F}^{(N-L)\times(N-L)})$.
Letting $\mathbf{S}_{\mathrm{F}}=\left[\begin{array}{cc}
\mathbf{I}_{L} & \mathbf{T}\\
\mathbf{0} & \mathbf{Z}
\end{array}\right]\mathbf{S}_{\mathrm{D}}$, we have
\begin{align*}
\mathbf{K}_{\mathrm{D}} & =\left[\mathbf{I}_{L}\,\big|\,\mathbf{T}\right]\mathbf{S}_{\mathrm{D}}
=\left[\mathbf{I}_{L}\,\big|\,\mathbf{0}\right]\left[\begin{array}{cc}
\mathbf{I}_{L} & \mathbf{T}\\
\mathbf{0} & \mathbf{Z}
\end{array}\right]\mathbf{S}_{\mathrm{D}}
=\left[\mathbf{I}_{L}\,\big|\,\mathbf{0}\right]\mathbf{S}_{\mathrm{F}},
\end{align*}
\begin{align*}
\mathbf{K}_{\mathrm{R}} & =\mathbf{S}_{\mathrm{D}}^{-1}\mathbf{S}_{\mathrm{E}}^{-1}
=\mathbf{S}_{\mathrm{F}}^{-1}\left[\begin{array}{cc}
\mathbf{I}_{L} & \mathbf{T}\\
\mathbf{0} & \mathbf{Z}
\end{array}\right]\mathbf{S}_{\mathrm{E}}^{-1}.
\end{align*}
We can then generate $\mathbf{K}_{\mathrm{R}},\mathbf{K}_{\mathrm{E}},\mathbf{K}_{\mathrm{D}}$
using $\mathbf{S}_{\mathrm{E}}\sim \mathrm{Unif}(\mathrm{FR}(\mathbb{F}^{N\times N}))$,
$\mathbf{Z}\sim \mathrm{Unif}(\mathrm{FR}(\mathbb{F}^{(N-L)\times(N-L)}))$,
and $\mathbf{S}_{\mathrm{F}}\sim \mathrm{Unif}(\mathrm{FR}(\mathbb{F}^{N\times N}))$
instead of $\mathbf{S}_{\mathrm{D}}$.
Since this way of generating
$\mathbf{K}_{\mathrm{R}},\mathbf{K}_{\mathrm{E}},\mathbf{K}_{\mathrm{D}}$
gives the correct distribution (uniform over $\mathcal{K}$) for any
fixed value of $\mathbf{Z}$, we may assume $\mathbf{Z}$ is generated
randomly as well.
So, we may also assume $\mathbf{K}_{\mathrm{R}},\mathbf{K}_{\mathrm{E}},\mathbf{K}_{\mathrm{D}}$
are generated via $\mathbf{S}_{\mathrm{E}}\sim \mathrm{Unif}(\mathrm{FR}(\mathbb{F}^{N\times N}))$,
$\mathbf{S}_{\mathrm{F}}\sim \mathrm{Unif}(\mathrm{FR}(\mathbb{F}^{N\times N}))$,
$\mathbf{Z}\sim \mathrm{Unif}(\mathrm{FR}(\mathbb{F}^{(N-L)\times(N-L)}))$,
$\mathbf{T}\sim \mathrm{Unif}(\mathbb{F}^{L\times(N-L)})$, and take
\begin{align}
\mathbf{K}_{\mathrm{E}} & \leftarrow\mathbf{S}_{\mathrm{E}}\left[\begin{array}{c}
\mathbf{I}_{L}\\
\mathbf{0}^{(N-L)\times L}
\end{array}\right],
\qquad
\nonumber \\
\mathbf{K}_{\mathrm{R}} 
& \leftarrow\mathbf{S}_{\mathrm{F}}^{-1}\left[\begin{array}{cc}
\mathbf{I}_{L} & \mathbf{T}\\
\mathbf{0} & \mathbf{Z}
\end{array}\right]\mathbf{S}_{\mathrm{E}}^{-1}, \qquad
\mathbf{K}_{\mathrm{D}} \leftarrow\left[\mathbf{I}_{L}\,\big|\,\mathbf{0}\right]\mathbf{S}_{\mathrm{F}}.\label{eq:keygen_alt}
\end{align}
Note that switching from \eqref{eq:keygen} to \eqref{eq:keygen_alt}
has no actual consequence to the setting because the joint distribution
of $\mathbf{K}_{\mathrm{R}},\mathbf{K}_{\mathrm{E}},\mathbf{K}_{\mathrm{D}}$
(which are the entities relevant to the actual setting) is preserved.
The matrices $\mathbf{S}_{\mathrm{E}},\mathbf{S}_{\mathrm{F}},\mathbf{Z},\mathbf{T}$
are abstract entities that only appear in the theoretical analysis.

We first consider the case where the leaker has $A=\{2\}$, and the
listener has $B=\{1,2\}$.
Assume the keys are generated according
to \eqref{eq:keygen_alt}.
Granting the leaker $\mathbf{S}_{\mathrm{F}}$,
we assume the leaker then produces $\mathbf{c} = f(\mathbf{S}_{\mathrm{F}}) \in \mathcal{C} = \mathbb{F}^{N}\backslash\{\mathbf{0}\}$.
For simplicity, we can assume $\mathbf{c}$ is a deterministic function
of $\mathbf{S}_{\mathrm{F}}$, since a randomized function provides
no benefit in 
maximizing the statistical distance \eqref{eq:setting_sd}.\footnote{If 
the leaker randomizes among several deterministic strategies, then the distribution of $p_{C',\mathbf{K}_{\mathrm{E}}}$
induced by the randomized strategy is a convex combination of the distributions induced by the deterministic strategies.
Since the statistical distance~\eqref{eq:setting_sd} is convex, 
the statistical distance given by the randomized strategy cannot be larger than the best deterministic strategy.}
After sanitization, it becomes 
\begin{align*}
\mathbf{c}' & =\mathbf{K}_{\mathrm{R}}f(\mathbf{S}_{\mathrm{F}})
=\mathbf{S}_{\mathrm{F}}^{-1}\left[\begin{array}{cc}
\mathbf{I}_{L} & \mathbf{T}\\
\mathbf{0} & \mathbf{Z}
\end{array}\right]\mathbf{S}_{\mathrm{E}}^{-1}f(\mathbf{S}_{\mathrm{F}}).
\end{align*}
Note that for any $\mathbf{v}\in\mathbb{F}^{N}$ with $\left[\mathbf{0}\,\big|\,\mathbf{I}_{N-L}\right]\mathbf{v}\neq\mathbf{0}$
(i.e., not all nonzero entries in $\mathbf{v}$ are at the first $L$
positions), $\left[\begin{array}{cc}
\mathbf{I}_{L} & \mathbf{T}\\
\mathbf{0} & \mathbf{Z}
\end{array}\right]\mathbf{v}$ is uniformly distributed over $\mathbb{F}^{L}\times(\mathbb{F}^{N-L}\backslash\{\mathbf{0}\})$
(i.e., the first~$L$ entries are uniformly distributed, and the last
$N-L$ entries are uniformly distributed among nonzero vectors).
So we have
\allowdisplaybreaks
\begin{align*}
 & \mathrm{SD}\left(\left[\begin{array}{cc}
\mathbf{I}_{L} & \mathbf{T}\\
\mathbf{0} & \mathbf{Z}
\end{array}\right]\mathbf{v},\,\mathrm{Unif}\left(\mathbb{F}^{N}\backslash\{\mathbf{0}\}\right)\right)\\
 & = 1-\frac{|\mathbb{F}^{L}\times(\mathbb{F}^{N-L}\backslash\{\mathbf{0}\})|}{|\mathbb{F}^{N}\backslash\{\mathbf{0}\}|}
 \; =\; \frac{q^{L}-1}{q^{N}-1},\text{~and}
\end{align*}
\begin{align*}
& 
\!\!\!\!
\mathrm{SD}\big(p_{\mathbf{c}',\mathbf{K}_{\mathrm{E}},\mathbf{K}_{\mathrm{D}}},\mathrm{Unif}\big((\mathbb{F}^{N}\backslash\{\mathbf{0}\})\times\mathrm{FR}(\mathbb{F}^{N\times L})
 \!\times\!
 \mathrm{FR}(\mathbb{F}^{L\times N})\!\big)\!\big)
 \\ \le &~ \mathrm{SD}\big(p_{\mathbf{c}',\mathbf{S}_{\mathrm{E}},\mathbf{S}_{\mathrm{F}}},\,\mathrm{Unif}\big((\mathbb{F}^{N}\backslash\{\mathbf{0}\})\times\mathrm{FR}(\mathbb{F}^{N\times N})^{2}\big)\big)
 \\ = &~ \mathbb{E}_{\mathbf{S}_{\mathrm{E}},\mathbf{S}_{\mathrm{F}}}\left[\mathrm{SD}\big(p_{\mathbf{c}'|\mathbf{S}_{\mathrm{E}},\mathbf{S}_{\mathrm{F}}},\,\mathrm{Unif}(\mathbb{F}^{N}\backslash\{\mathbf{0}\})\big)\right]
 \\ \le &~ \mathbb{P}\left(\left[\mathbf{0}\big|\mathbf{I}_{N-L}\right]\mathbf{S}_{\mathrm{E}}^{-1}f(\mathbf{S}_{\mathrm{F}})\! =\! \mathbf{0}\right) + \mathbb{P}\left(\left[\mathbf{0}\big|\mathbf{I}_{N-L}\right]\mathbf{S}_{\mathrm{E}}^{-1}f(\mathbf{S}_{\mathrm{F}})\! \neq\! \mathbf{0}\right)
 \\ &\!\!\!\cdot \mathbb{E}\left[\mathrm{SD}\big(p_{\mathbf{c}'|\mathbf{S}_{\mathrm{E}},\mathbf{S}_{\mathrm{F}}}\!,\!\mathrm{Unif}(\mathbb{F}^{N}\!\backslash\{\mathbf{0}\})\big)\, \big| \, \left[\mathbf{0}\big|\mathbf{I}_{N-L}\right]\mathbf{S}_{\mathrm{E}}^{-1}f(\mathbf{S}_{\mathrm{F}})\!\neq\! \mathbf{0}\right]
 \\ \le &~ \mathbb{P}\left(\left[\mathbf{0}\,\big|\,\mathbf{I}_{N-L}\right]\mathbf{S}_{\mathrm{E}}^{-1}f(\mathbf{S}_{\mathrm{F}})=\mathbf{0}\right)+\frac{q^{L}-1}{q^{N}-1}
 \\ = &~ \frac{2(q^{L}-1)}{q^{N}-1} 
 \\ \le& ~ 2q^{-(N-L)}.
\end{align*}

\ifshortver
\else
We then consider the case where the leaker has $A=\{1,2\}$, and the
listener has $B=\{1\}$. Before we prove this case, we consider another
alternative method of generating $(\mathbf{K}_{\mathrm{R}},\mathbf{K}_{\mathrm{E}},\mathbf{K}_{\mathrm{D}})$
uniformly from $\mathcal{K}$ based on \eqref{eq:keygen_alt}. Since
$(\mathbf{K}_{\mathrm{R}},\mathbf{K}_{\mathrm{E}},\mathbf{K}_{\mathrm{D}})\in\mathcal{K}$
if and only if $(\mathbf{K}_{\mathrm{R}}^{T},\mathbf{K}_{\mathrm{D}}^{T},\mathbf{K}_{\mathrm{E}}^{T})\in\mathcal{K}$,
by taking the transpose of \eqref{eq:keygen_alt}, we may also assume
$\mathbf{K}_{\mathrm{R}},\mathbf{K}_{\mathrm{E}},\mathbf{K}_{\mathrm{D}}$
are generated via $\mathbf{S}_{\mathrm{E}}\sim\mathrm{Unif}(\mathrm{FR}(\mathbb{F}^{N\times N}))$,
$\mathbf{S}_{\mathrm{G}}\sim\mathrm{Unif}(\mathrm{FR}(\mathbb{F}^{N\times N}))$,
$\mathbf{Z}\sim\mathrm{Unif}(\mathrm{FR}(\mathbb{F}^{(N-L)\times(N-L)}))$,
$\mathbf{U}\sim\mathrm{Unif}(\mathbb{F}^{(N-L)\times L})$, and take
\begin{align}
\mathbf{K}_{\mathrm{E}} & \leftarrow\mathbf{S}_{\mathrm{E}}\left[\begin{array}{c}
\mathbf{I}_{L}\\
\mathbf{0}^{(N-L)\times L}
\end{array}\right],\qquad\nonumber \\
\mathbf{K}_{\mathrm{R}} & \leftarrow\mathbf{S}_{\mathrm{G}}^{-1}\left[\begin{array}{cc}
\mathbf{I}_{L} & \mathbf{0}\\
\mathbf{U} & \mathbf{Z}
\end{array}\right]\mathbf{S}_{\mathrm{E}}^{-1},\nonumber \\
\mathbf{K}_{\mathrm{D}} & \leftarrow\left[\mathbf{I}_{L}\,\big|\,\mathbf{0}\right]\mathbf{S}_{\mathrm{G}}.\label{eq:keygen_alt2}
\end{align}
Assume the keys are generated according to \eqref{eq:keygen_alt2},
and we further allow the leaker to access $\mathbf{S}_{\mathrm{E}},\mathbf{S}_{\mathrm{G}}$.
Assume the leaker produces $\mathbf{c}=f(\mathbf{S}_{\mathrm{E}},\mathbf{S}_{\mathrm{G}})$.
After sanitization, it becomes 
\begin{align}
\mathbf{c}' & =\mathbf{K}_{\mathrm{R}}f(\mathbf{S}_{\mathrm{E}},\mathbf{S}_{\mathrm{G}})\nonumber \\
 & =\mathbf{S}_{\mathrm{G}}^{-1}\left[\begin{array}{cc}
\mathbf{I}_{L} & \mathbf{0}\\
\mathbf{U} & \mathbf{Z}
\end{array}\right]\mathbf{S}_{\mathrm{E}}^{-1}f(\mathbf{S}_{\mathrm{E}},\mathbf{S}_{\mathrm{G}})\nonumber \\
 & =\mathbf{S}_{\mathrm{G}}^{-1}\left[\begin{array}{cc}
\mathbf{I}_{L} & \mathbf{0}\\
\mathbf{U} & \mathbf{Z}
\end{array}\right]\left[\begin{array}{c}
\mathbf{x}\\
\mathbf{y}
\end{array}\right]\nonumber \\
 & =\mathbf{S}_{\mathrm{G}}^{-1}\left[\begin{array}{c}
\mathbf{x}\\
\mathbf{U}\mathbf{x}+\mathbf{Z}\mathbf{y}
\end{array}\right],\label{eq:nwfp_cp}
\end{align}
where we let $\left[\begin{array}{c}
\mathbf{x}\\
\mathbf{y}
\end{array}\right]=\left[\begin{array}{c}
\mathbf{x}(\mathbf{S}_{\mathrm{E}},\mathbf{S}_{\mathrm{G}})\\
\mathbf{y}(\mathbf{S}_{\mathrm{E}},\mathbf{S}_{\mathrm{G}})
\end{array}\right]=\mathbf{S}_{\mathrm{E}}^{-1}f(\mathbf{S}_{\mathrm{E}},\mathbf{S}_{\mathrm{G}})$, $\mathbf{x}\in\mathbb{F}^{L}$, $\mathbf{y}\in\mathbb{F}^{N-L}$,
$(\mathbf{x},\mathbf{y})\neq(\mathbf{0},\mathbf{0})$ (while $\mathbf{x}(\mathbf{S}_{\mathrm{E}},\mathbf{S}_{\mathrm{G}})$
is a function of $\mathbf{S}_{\mathrm{E}},\mathbf{S}_{\mathrm{G}}$,
we omit ``$(\mathbf{S}_{\mathrm{E}},\mathbf{S}_{\mathrm{G}})$''
and only write $\mathbf{x}$ for brevity). It is clear that $\mathbf{c}'$
is conditionally independent of $\mathbf{S}_{\mathrm{E}}$ given $(\mathbf{x},\mathbf{y})$.
Conditional on any pair $(\mathbf{x},\mathbf{y})$ where $\mathbf{x}\neq0$,
we have the conditional distribution 
\begin{equation}
\left[\begin{array}{c}
\mathbf{x}\\
\mathbf{U}\mathbf{x}+\mathbf{Z}\mathbf{y}
\end{array}\right]\sim\mathrm{Unif}\left(\left\{ \left[\begin{array}{c}
\mathbf{x}\\
\mathbf{w}
\end{array}\right]:\,\mathbf{w}\in\mathbb{F}^{N-L}\right\} \right),\label{eq:nwfp_y0}
\end{equation}
i.e., the vector on the left-hand side is uniformly distributed over
the set of vectors with the first $L$ entries agreeing with $\mathbf{x}$.
Conditional on any pair $(\mathbf{x},\mathbf{y})$ where $\mathbf{x}=0$
(which forces $\mathbf{y}\neq\mathbf{0}$ since $\mathbf{S}_{\mathrm{E}}^{-1}f(\mathbf{S}_{\mathrm{E}},\mathbf{S}_{\mathrm{G}})\neq\mathbf{0}$),
we have the conditional distribution 
\begin{align}
\left[\begin{array}{c}
\mathbf{x}\\
\mathbf{U}\mathbf{x}+\mathbf{Z}\mathbf{y}
\end{array}\right] & \sim\mathrm{Unif}\left(\left\{ \left[\begin{array}{c}
\mathbf{0}\\
\mathbf{w}
\end{array}\right]:\,\mathbf{w}\in\mathbb{F}^{N-L}\backslash\{\mathbf{0}\}\right\} \right).\label{eq:nwfp_yn0}
\end{align}
We can see that \eqref{eq:nwfp_y0} almost also holds when $\mathbf{x}=0$,
with the only difference being whether $\mathbf{w}$ can be $\mathbf{0}$.
The statistical distance between \eqref{eq:nwfp_y0} (where we substitute
$\mathbf{x}=0$ ignoring the requirement that $\mathbf{x}\neq0$)
and \eqref{eq:nwfp_yn0} is upper-bounded by $q^{-(N-L)}$. Therefore,
we will modify the conditional distribution of $\mathbf{c}'$ given
$(\mathbf{S}_{\mathrm{G}},\mathbf{x},\mathbf{y})$, and assume that
\begin{equation}
\mathbf{c}'=\mathbf{S}_{\mathrm{G}}^{-1}\left[\begin{array}{c}
\mathbf{x}\\
\mathbf{w}
\end{array}\right]\label{eq:nwfp_mod}
\end{equation}
instead of \eqref{eq:nwfp_cp}, where $\mathbf{w}\sim\mathrm{Unif}(\mathbb{F}^{N-L})$
independent of $(\mathbf{S}_{\mathrm{G}},\mathbf{x},\mathbf{y})$.
The modified conditional distribution of $\mathbf{c}'$ given $(\mathbf{S}_{\mathrm{G}},\mathbf{x},\mathbf{y})$
is within a statistical distance $q^{-(N-L)}$ from the original conditional
distribution.

We now assume $\mathbf{c}'$ is generated according to the modified
distribution \eqref{eq:nwfp_mod}. Define 
\begin{align*}
\mathcal{V} & =\mathcal{V}(\mathbf{S}_{\mathrm{G}})=\left\{ \mathbf{S}_{\mathrm{G}}^{-1}\left[\begin{array}{c}
\mathbf{0}\\
\mathbf{w}
\end{array}\right]:\,\mathbf{w}\in\mathbb{F}^{N-L}\right\} ,\\
\mathbf{g} & =\mathbf{g}(\mathbf{S}_{\mathrm{E}},\mathbf{S}_{\mathrm{G}})=\mathbf{S}_{\mathrm{G}}^{-1}\left[\begin{array}{c}
\mathbf{x}\\
\mathbf{0}
\end{array}\right].
\end{align*}
Note that $\mathcal{V}$ is a uniformly randomly chosen $(N-L)$-dimensional
subspace of $\mathbb{F}^{N}$. From \eqref{eq:nwfp_mod}, $\mathbf{c}'$
is uniformly distributed over 
\[
\mathcal{V}+\mathbf{g}=\left\{ \mathbf{v}+\mathbf{g}:\,\mathbf{v}\in\mathcal{V}\right\}.
\]
We want to argue that $\mathbf{c}'$ is close to uniformly distributed
conditional on any value of $\mathbf{S}_{\mathrm{E}}$. To this end,
we generate another copy $(\bar{\mathcal{V}},\bar{\mathbf{g}},\bar{\mathbf{c}}')$
independent of $(\mathcal{V},\mathbf{g},\mathbf{c}')$ and with the
same distribution conditional on $\mathbf{S}_{\mathrm{E}}$. We have
\begin{align}
 & \mathbb{P}\left(\bar{\mathbf{c}}'=\mathbf{c}'\,\big|\,\mathbf{S}_{\mathrm{E}}\right)\nonumber \\
 & =\mathbb{E}_{\mathcal{V},\bar{\mathcal{V}},\mathbf{g},\bar{\mathbf{g}}}\left[\mathbb{P}\left(\bar{\mathbf{c}}'=\mathbf{c}'\,\big|\,\mathcal{V},\bar{\mathcal{V}},\mathbf{g},\bar{\mathbf{g}}\right)\,\big|\,\mathbf{S}_{\mathrm{E}}\right]\nonumber \\
 & =\mathbb{E}\left[q^{-2(N-L)}\left|(\mathcal{V}+\mathbf{g})\cap(\bar{\mathcal{V}}+\bar{\mathbf{g}})\right|\,\big|\,\mathbf{S}_{\mathrm{E}}\right]\nonumber \\
 & =\mathbb{E}\left[q^{-2(N-L)}\left|\mathcal{V}\cap(\bar{\mathcal{V}}+(\bar{\mathbf{g}}-\mathbf{g}))\right|\,\big|\,\mathbf{S}_{\mathrm{E}}\right]\nonumber \\
 & \stackrel{(a)}{\le}\mathbb{E}\left[q^{-2(N-L)}\left|\mathcal{V}\cap\bar{\mathcal{V}}\right|\,\big|\,\mathbf{S}_{\mathrm{E}}\right]\nonumber \\
 & =q^{-2(N-L)}\sum_{\mathbf{v}\in\mathbb{F}^{N}}\mathbb{P}(\mathbf{v}\in\mathcal{V}\,|\,\mathbf{S}_{\mathrm{E}})\mathbb{P}(\mathbf{v}\in\bar{\mathcal{V}}\,|\,\mathbf{S}_{\mathrm{E}})\nonumber \\
 & =q^{-2(N-L)}\left(1+(q^{N}-1)\left(\frac{q^{N-L}-1}{q^{N}-1}\right)^{2}\right)\nonumber \\
 & =q^{-2(N-L)}\left(1+\frac{(q^{N-L}-1)^{2}}{q^{N}-1}\right)\nonumber \\
 & \le q^{-2(N-L)}\left(1+\frac{q^{N-L}(q^{N-L}-1)}{q^{N}}\right)\nonumber \\
 & \le q^{-2(N-L)}\left(1+q^{N-2L}\right)\nonumber \\
 & =q^{-2(N-L)}+q^{-N},\label{eq:nwfp_coin}
\end{align}
when $L\le N/2$, where (a) is because $\mathcal{V}\cap(\bar{\mathcal{V}}+\mathbf{g})$
is either empty or a coset of the subspace $\mathcal{V}\cap\bar{\mathcal{V}}$
in the vector space $\mathcal{V}$ (since if $\mathbf{x}\in\mathcal{V}\cap(\bar{\mathcal{V}}+\mathbf{g})$,
then $\mathbf{y}\in\mathcal{V}\cap(\bar{\mathcal{V}}+\mathbf{g})$
if and only if $\mathbf{y}-\mathbf{x}\in\mathcal{V}\cap\bar{\mathcal{V}}$,
and hence $\mathcal{V}\cap(\bar{\mathcal{V}}+\mathbf{g})=(\mathcal{V}\cap\bar{\mathcal{V}})+\mathbf{x}$),
each coset of the same subspace has the same size, and the particular
coset $\mathcal{V}\cap\bar{\mathcal{V}}$ is nonempty (since $\mathbf{0}\in\mathcal{V}\cap\bar{\mathcal{V}}$),
implying that $\mathcal{V}\cap\bar{\mathcal{V}}$ is one of the largest
possible $\mathcal{V}\cap(\bar{\mathcal{V}}+\mathbf{g})$. Since for
any probability mass function $p(x)$ for $x\in[n]$, we have 
\begin{align*}
\sum_{x=1}^{n}(p(x))^{2} & =\sum_{x}\left(p(x)-\frac{1}{n}\right)^{2}+\frac{1}{n}\\
 & \ge n\left(\frac{1}{n}\sum_{x}\left|p(x)-\frac{1}{n}\right|\right)^{2}+\frac{1}{n}\\
 & =\frac{4}{n}\left(\mathrm{SD}(p,\mathrm{Unif}([n]))\right)^{2}+\frac{1}{n},
\end{align*}
we know that by \eqref{eq:nwfp_coin}, 
\begin{align*}
 & q^{-2(N-L)}+q^{-N}\\
 & \ge\mathbb{P}\left(\bar{\mathbf{c}}'=\mathbf{c}'\,\big|\,\mathbf{S}_{\mathrm{E}}\right)\\
 & =\sum_{\mathbf{v}\in\mathbb{F}^{N}}\left(\mathbb{P}\left(\mathbf{c}'=\mathbf{v}\,\big|\,\mathbf{S}_{\mathrm{E}}\right)\right)^{2}\\
 & \ge\frac{4}{q^{N}}\left(\mathrm{SD}(p_{\mathbf{c}'|\mathbf{S}_{\mathrm{E}}},\,\mathrm{Unif}(\mathbb{F}^{N}))\right)^{2}+\frac{1}{q^{N}}.
\end{align*}
Rearranging the terms, 
\begin{align*}
\mathrm{SD}(p_{\mathbf{c}'|\mathbf{S}_{\mathrm{E}}},\,\mathrm{Unif}(\mathbb{F}^{N})) & \le\frac{1}{2}\sqrt{q^{-(N-2L)}}\\
 & =\frac{1}{2}q^{-(N/2-L)}
\end{align*}
since $N\ge2$, $L\ge1$, $q\ge2$. Hence, if $\mathbf{c}'$ is generated
according to the modified distribution \eqref{eq:nwfp_mod}, we can
guarantee that 
\begin{align*}
 & \mathrm{SD}(p_{\mathbf{c}',\mathbf{S}_{\mathrm{E}}},\,\mathrm{Unif}((\mathbb{F}^{N}\backslash\{\mathbf{0}\})\times\mathrm{FR}(\mathbb{F}^{N\times N})))\\
 & =\mathbb{E}_{\mathbf{S}_{\mathrm{E}}}\left[\mathrm{SD}(p_{\mathbf{c}'|\mathbf{S}_{\mathrm{E}}},\,\mathrm{Unif}(\mathbb{F}^{N}\backslash\{\mathbf{0}\}))\right]\\
 & \le\mathbb{E}_{\mathbf{S}_{\mathrm{E}}}\left[\mathrm{SD}(p_{\mathbf{c}'|\mathbf{S}_{\mathrm{E}}},\,\mathrm{Unif}(\mathbb{F}^{N}))+\mathrm{SD}(\mathrm{Unif}(\mathbb{F}^{N}),\,\mathrm{Unif}(\mathbb{F}^{N}\backslash\{\mathbf{0}\}))\right]\\
 & \le\frac{1}{2}q^{-(N/2-L)}+q^{-N}.
\end{align*}
Therefore, for the original distribution \eqref{eq:nwfp_cp} of $\mathbf{c}'$ (which is within a statistical distance $q^{-(N-L)}$ from \eqref{eq:nwfp_mod}),
\begin{align*}
 & \mathrm{SD}(p_{\mathbf{c}',\mathbf{S}_{\mathrm{E}}},\,\mathrm{Unif}((\mathbb{F}^{N}\backslash\{\mathbf{0}\})\times\mathrm{FR}(\mathbb{F}^{N\times N})))\\
 & \le\frac{1}{2}q^{-(N/2-L)}+q^{-N}+q^{-(N-L)}\\
 & \le\frac{1}{2}q^{-(N/2-L)}+\frac{1}{4}q^{-(N/2-L)}+\frac{1}{2}q^{-(N/2-L)}\\
 & =\frac{5}{4}q^{-(N/2-L)}
\end{align*}
since $N\ge2$, $L\ge1$, $q\ge2$. Combining the two cases of the
no-write rule, we have the distance parameter 
\begin{align*}
\epsilon & =\max\left\{ 2q^{-(N-L)},\,\frac{5}{4}q^{-(N/2-L)}\right\} \\
 & \le\max\left\{ q^{-(N/2-L)},\,\frac{5}{4}q^{-(N/2-L)}\right\} \\
 & =\frac{5}{4}q^{-(N/2-L)}.
\end{align*}
\fi\end{IEEEproof}

\medskip
\section{Our Final Scheme for General Access Policies\label{sec:general}}

To construct a scheme for a general access policy $E$, we can simply
run $|E|$ copies of the random matrix scheme for a single sender-receiver
pair.
For each $(i,j)\in E$, run the key generation of the random
matrix scheme independently to produce $\mathbf{K}_{\mathrm{R},i,j},\mathbf{K}_{\mathrm{E},i,j},\mathbf{K}_{\mathrm{D},i,j}$.
The sanitizer has the key $K_0 = (\mathbf{K}_{\mathrm{R},i,j})_{(i,j)\in E}$.
Party $a$ has the key $K_{a}=((\mathbf{K}_{\mathrm{E},a,j})_{j:(a,j)\in E},(\mathbf{K}_{\mathrm{D},i,a})_{i:(i,a)\in E})$.
If party $a$ wants to send~$\mathbf{m}$ to party $b$ legitimately
($(a,b)\in E$), party $a$ produces the ciphertext $C=(\mathbf{c}_{i,j})_{(i,j)\in E}\in(\mathbb{F}^{L}\backslash\{\mathbf{0}\})^{|E|}$,
where
\[
\mathbf{c}_{i,j}\begin{cases}
=\mathbf{K}_{\mathrm{E},a,b}\mathbf{m} & \mathrm{if}\;(i,j)=(a,b)\\
\sim\mathrm{Unif}(\mathbb{F}^{L}\backslash\{\mathbf{0}\}) & \mathrm{if}\;(i,j)\neq(a,b).
\end{cases}
\]
Upon receiving $(\mathbf{c}_{i,j})_{(i,j)\in E}\in(\mathbb{F}^{L}\backslash\{\mathbf{0}\})^{|E|}$,
the sanitizer outputs $(\mathbf{c}'_{i,j})_{(i,j)\in E}=(\mathbf{K}_{\mathrm{R},i,j}\mathbf{c}_{i,j})_{(i,j)\in E}\in(\mathbb{F}^{L}\backslash\{\mathbf{0}\})^{|E|}$.
Party $b$ can recover the message from party $a$ by $\hat{\mathbf{m}}=\mathbf{K}_{\mathrm{D},a,b}\mathbf{c}'_{a,b}$.
We can prove the security of this scheme using Theorem \ref{thm:sec}, with a penalty on the distance parameter that scales linearly with the number of sender-receiver pairs $|E|$.

\medskip
\begin{thm}
\label{thm:general}
The random matrix protocol for general $E$ attains a distance parameter
\ifshortver
\else
\[
\epsilon=2|E|q^{-(N/2-L)}.
\]
\fi
\end{thm}
\medskip

\ifshortver
\else
\begin{IEEEproof}
Correctness and the perfect no-read rule follow directly from Theorem~\ref{thm:sec}.
To prove the no-write rule,
let $\epsilon_{0}=2q^{-(N/2-L)}$.
Let the edges be $E=\{e_{1},e_{2},\ldots,e_{|E|}\}$, where $e_{i}\in[n]^{2}$.
For brevity, we write $\mathbf{c}_{i}=\mathbf{c}_{e_{i,1},e_{i,2}}$
where $e_{i}=(e_{i,1},e_{i,2})$, i.e., $e_i$ is the edge from party $e_{i,1}$ to party $e_{i,2}$.
Write $\mathbf{K}_{\mathrm{R},i},\mathbf{K}_{\mathrm{E},i},\ldots$
similarly.
Write $\mathbf{K}_{\mathrm{O},i}=(\mathbf{K}_{\mathrm{R},i},\mathbf{K}_{\mathrm{E},i},\mathbf{K}_{\mathrm{D},i})$.
Consider a leaker with keys $K_{A}$ and a listener with keys $K_{B}$,
$A, B \subseteq [n]$.
Write $\mathbf{K}_{B,i}$ for the collection of
keys that the listener has among $\mathbf{K}_{\mathrm{E},i},\mathbf{K}_{\mathrm{D},i}$
(recall that the listener has $\mathbf{K}_{\mathrm{E},i}$ if $e_{i,1}\in B$,
and has $\mathbf{K}_{\mathrm{D},i}$ if $e_{i,2}\in B$).
Assume the leaker produces $C=(\mathbf{c}_{i})_{i\in[|E|]}$ that is a function of $K_{A}$ (a randomized attack cannot increase the statistical distance;
see the proof of Theorem \ref{thm:sec}).
The sanitizer outputs $\mathbf{c}'_{i}=\mathbf{K}_{\mathrm{R},i}\mathbf{c}_{i}$.
For each $i\in[|E|]$, by Theorem \ref{thm:sec}, we have
\begin{align*}
 & \mathrm{SD}\big(p_{\mathbf{c}'_{i},\mathbf{K}_{B,i}},\,\mathrm{Unif}(\mathbb{F}^{L}\backslash\{\mathbf{0}\})\times p_{\mathbf{K}_{B,i}}\big)\le\epsilon_{0}.
\end{align*}
This holds since the keys $\mathbf{K}_{\mathrm{R},i},\mathbf{K}_{\mathrm{E},i},\mathbf{K}_{\mathrm{D},i}$
for different slots~$i$ are independent, so the leaker knowing keys at a slot other than $i$ will not help increase the statistical distance between $\mathbf{c}'_{i}$ and the uniform distribution.
It is left to combine these bounds for different $i$.
To this end, we employ a coupling strategy.
We will construct random
variables $\bar{\mathbf{c}}'_{i}\in\mathbb{F}^{L}\backslash\{\mathbf{0}\}$
recursively, satisfying the following ``recursive independence condition'':
\[
\bar{\mathbf{c}}'_{1},\ldots,\bar{\mathbf{c}}'_{i},\mathbf{K}_{B,1},\ldots,\mathbf{K}_{B,i},\mathbf{K}_{\mathrm{O},i+1},\ldots,\mathbf{K}_{\mathrm{O},|E|}
\]
are mutually independent.
Consider $i\in[|E|]$.
Assume that the recursive independence condition holds for \mbox{$i-1$}, i.e., 
\[
\bar{\mathbf{c}}'_{1},\ldots,\bar{\mathbf{c}}'_{i-1},\mathbf{K}_{B,1},\ldots,\mathbf{K}_{B,i-1},\mathbf{K}_{\mathrm{O},i},\ldots,\mathbf{K}_{\mathrm{O},|E|}
\]
are mutually independent.
We construct $\bar{\mathbf{c}}'_{i}$
satisfying the recursive independence assumption below.
Applying the no-write
rule guarantee in Theorem \ref{thm:sec} conditional on $\bar{\mathbf{c}}'_{<i},\mathbf{K}_{B,<i},\mathbf{K}_{\mathrm{O},>i}$,
where we write $\bar{\mathbf{c}}'_{<i}=(\bar{\mathbf{c}}'_{1},\ldots,\bar{\mathbf{c}}'_{i-1})$
(i.e., consider~the random matrix scheme restricted to the slot
$\mathbf{c}_{i}$, ignoring other slots $(\mathbf{c}_{i'})_{i'\neq i}$,
while conditioning on the values of $\bar{\mathbf{c}}'_{<i},\mathbf{K}_{B,<i},\mathbf{K}_{\mathrm{O},>i}$,
which will not affect the distribution of $\mathbf{K}_{\mathrm{O},i}$
assuming the recursive independence condition,
so the scheme
is still valid under the conditional distribution),
\begin{align*}
 & \mathrm{SD}\big(p_{\mathbf{c}'_{i},\mathbf{K}_{B,i}|\bar{\mathbf{c}}'_{<i},\mathbf{K}_{B,<i},\mathbf{K}_{\mathrm{O},>i}},\mathrm{Unif}(\mathbb{F}^{L}\backslash\{\mathbf{0}\})\times p_{\mathbf{K}_{B,i}}\big)\le\epsilon_{0}.
\end{align*}
By the coupling property of statistical distance, 
we can construct
a random variable $\bar{\mathbf{c}}'_{i}$ (conditionally independent~of all previously defined random variables given $\mathbf{c}'_{i},\bar{\mathbf{c}}'_{<i},\mathbf{K}_{B,\le i},\mathbf{K}_{\mathrm{O},>i}$) that is uniform over $\mathbb{F}^{L}\backslash\{\mathbf{0}\}$
and independent of $(\mathbf{K}_{B,i},\bar{\mathbf{c}}'_{<i},\mathbf{K}_{B,<i},\mathbf{K}_{\mathrm{O},>i})$,
such that $\mathbb{P}(\mathbf{c}'_{i}\neq\bar{\mathbf{c}}'_{i})\le\epsilon_{0}$.
With the recursive independence condition for $i\!-\!1$,
this
gives the recursive independence condition for~$i$.

Repeating this construction, we can have mutually independent random variables $\bar{\mathbf{c}}'_{1},\ldots,\bar{\mathbf{c}}'_{|E|}$,
each uniform over $\mathbb{F}^{L} \backslash\{\mathbf{0}\}$, and mutually
independent of $\mathbf{K}_{B,1},\ldots,\mathbf{K}_{B,|E|}$, i.e.,
\begin{align*}
 & p_{\bar{\mathbf{c}}'_{1},\ldots,\bar{\mathbf{c}}'_{|E|},\mathbf{K}_{B,1},\ldots,\mathbf{K}_{B,|E|}}\\
 & =\mathrm{Unif}\big((\mathbb{F}^{L}\backslash\{\mathbf{0}\})^{|E|}\big)\times p_{\mathbf{K}_{B,1},\ldots,\mathbf{K}_{B,|E|}}.
\end{align*}
The desired bound on the statistical distance in the no-write rule
follows from $\mathbb{P}(\mathbf{c}'_{i}\neq\bar{\mathbf{c}}'_{i})\le\epsilon_{0}$,
which gives
$\mathbb{P}((\bar{\mathbf{c}}'_{1},\ldots,\bar{\mathbf{c}}'_{|E|})\neq(\mathbf{c}'_{1},\ldots,\mathbf{c}'_{|E|}))\le|E|\epsilon_{0}$.
\end{IEEEproof}
\fi

\medskip
\section{Concluding Remarks}
Access control encryption (ACE) is a useful cryptographic primitive that enforces information flow, which addresses the root cause of recurrent data breaches by preventing illegitimate information leakage.
The distinctive feature of ACE is that a sanitizer can enforce access control without learning the plaintext data, the sender and the receiver identities, and even whether the communication is a legitimate one allowed by the access control policy.
Many ACE schemes have been proposed; however, they are only secure against computationally-bounded adversaries.
It is unclear how to construct ACE without using existing public-key techniques.

In this paper, we have presented the first unconditionally secure ACE scheme, with a novel construction where a triple of random matrices (under the constraint that their product is the identity matrix) are used as secret keys.
We hope our result inspires further information theory research for the atypical communication model underlying ACE.

A shortcoming of Theorem~\ref{thm:general} is that the ciphertext length needs to scale linearly with the number of sender-receiver pairs~$|E|$.
We leave it for future studies to determine whether this factor can be reduced to $O(n)$~\cite{damgaard2016access} or, ultimately, $O(1)$~\cite{wang2021cross} as in computationally secure ACE schemes while retaining the unconditional security guarantee.
The optimal scaling of the key size with respect to the message size would be another interesting problem for future studies.

\ifshortver
\else
\section{Acknowledgement}
\ackStmt
\fi

\bibliographystyle{IEEEtran}
\bibliography{ace}





\end{document}